\documentclass{appolb}

\usepackage{wrapfig, graphicx, color, subfigure, slashed, listings, fancyhdr, amsmath, amsthm, amssymb, bbm, dsfont}

\begin{document}

\title{Infrared behaviour of propagators and running coupling in the conformal window of QCD\thanks{Presented at Excited QCD 2014, 
$2^{nd}$ -- $8^{th}$ February 2014, Bjelasnica Mountain,  Sarajevo, Bosnia-Herzegovina}}

\author{
Markus Hopfer, Reinhard Alkofer
\address{Institut f\"ur Physik, Karl-Franzens Universit\"at Graz\\ Universit\"atsplatz 5, 8010 Graz, Austria}
\\[0.5cm]
Christian S. Fischer
\address{Institut f\"ur Theoretische Physik, Justus-Liebig Universit\"at Giessen,\\Heinrich-Buff-Ring 16, 35392 Giessen, Germany}
}

\maketitle

\begin{abstract}
Using the Dyson-Schwinger approach we investigate Landau gauge QCD with a relatively large number of chiral quark flavours.
A self-consistent treatment on the propagator level enables us to study unquenching effects 
via the quark loop diagram in the gluon equation. 
Above the critical number of fermion flavours the non-perturbative running coupling develops a plateau over a wide  
momentum range. Correspondingly, the propagators follow a power law behaviour in this momentum range 
indicating conformal behaviour.  
Our value $N_f^{crit}=4.5$ is strongly sensitive to the details of the  
quark-gluon vertex calling for more detailed investigations in future studies.
\end{abstract}


\section{Introduction}

Walking technicolor models have been introduced to overcome the phenomenological difficulties faced by 
the early technicolor formulations \cite{Weinberg:1979bn}, see \cite{Sannino:2009za} for a recent review.
These models exhibit an approximate scale invariance over a wide energy range as well as a 
proximity to an infrared fixed point, where the gauge coupling is slowly running, or walking.
Asymptotically free gauge theories can be utilized to mimic these properties,
where it is expected that QCD with a large number of chiral fermion flavours is a viable candidate.

By linking the Green functions of a quantum field theory 
the Dyson-Schwinger framework offers an appropriate non-perturbative tool to explore a given theory over 
all energies ranging from the deep infrared to the perturbative regime.
Since DSEs constitute an infinite set of coupled integral equations
carefully chosen truncations have to be applied in order to treat the equations numerically.
Therefore, a comparison with other non-perturbative methods is inevitable at some point
in order to fine-tune the truncation and to minimize errors induced by it.
Once an appropriate truncation scheme is established the Dyson-Schwinger framework is a reliable 
and robust tool to explore the theory.

\section{The System of Coupled Dyson-Schwinger Equations}
\label{sec:intro_coupled_system}

\begin{wrapfigure}[8]{r}{0.55\columnwidth}
\center
\vspace{-0.6cm}
\includegraphics[width=0.55\columnwidth]{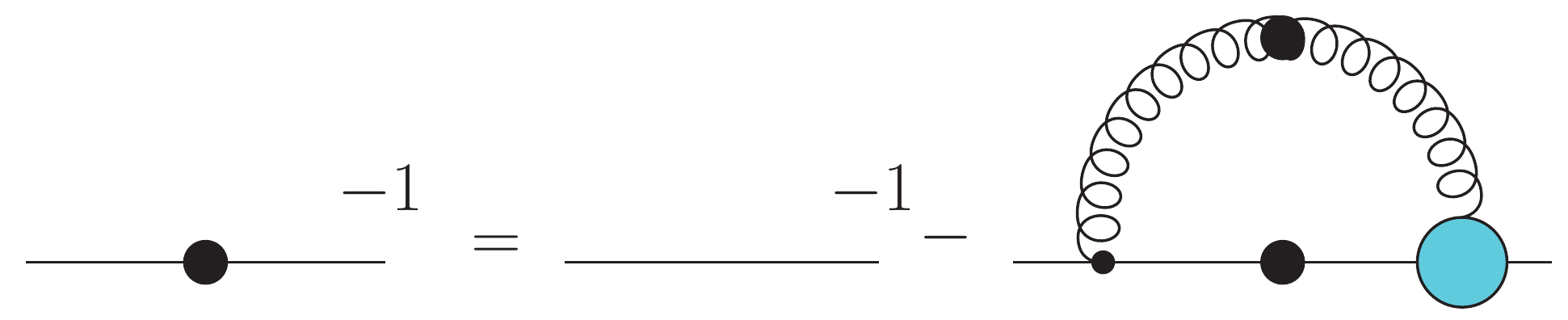} \\
\caption{The DSE for the quark propagator. All internal propagators are dressed. Coloured blobs denote full vertices.}
\label{fig:quark_DSE}
\end{wrapfigure}

The central object in the following investigation is the quark propagator DSE depicted in Fig.~\ref{fig:quark_DSE}.
Intimately connected is the dressed gluon propagator indicated by the wiggly line.
By increasing the fermion flavours, back-coupling effects of quark degrees of freedom on the Yang-Mills sector 
become important and simple model descriptions of the gluon propagator without detailed knowledge
of its flavour dependence will prove to be insufficient.
In particular, a naive extrapolation of QCD results to larger flavour numbers 
as done in Ref.~\cite{Bashir:2013zha} seems to be questionable.
Thus, a self-consistent incorporation of the corresponding gluon DSE becomes mandatory.
In the following we outline the coupled system of DSEs and refer to Ref.~\cite{Hopfer:2014zna} 
for details on the self-consistent treatment.

\subsection{The Coupled System}
\label{sec:coupled_system}
The renormalized DSE for the quark propagator is given by\footnote{We follow the conventions 
and notation of Ref.~\cite{Fischer:2003rp}.}
 \begin{equation}
 \label{eq:quark_DSE}
 S^{-1}(p) = Z_2 S_0^{-1}(p) + g^2Z_{1F}C_F\int\frac{d^4q}{(2\pi)^4} 
 \gamma^\mu S(q)\varGamma^\nu(q,p;k)D^{\mu\nu}(k).
 \end{equation}
Here, $Z_2$ and $Z_{1F}$ are the renormalization constants for the quark wave function and the quark-gluon vertex, respectively. 
The colour trace yields a factor of $C_F=(N_c^2-1)/(2N_c)$ and the gluon momentum is defined via $k_\mu=p_\mu-q_\mu$.
The full quark propagator is given by $S^{-1}(p) = -i\slashed p A(p^2,\mu^2)+B(p^2,\mu^2)$,
where the dressing functions $A$ and $B$ implicitly depend on the renormalization scale $\mu$.
The quark mass function is defined via $M(p^2)=B(p^2,\mu^2)/A(p^2,\mu^2)$ and is a renormalization scale independent quantity.

\begin{figure*}[!ht]
\center
\includegraphics[width=\columnwidth]{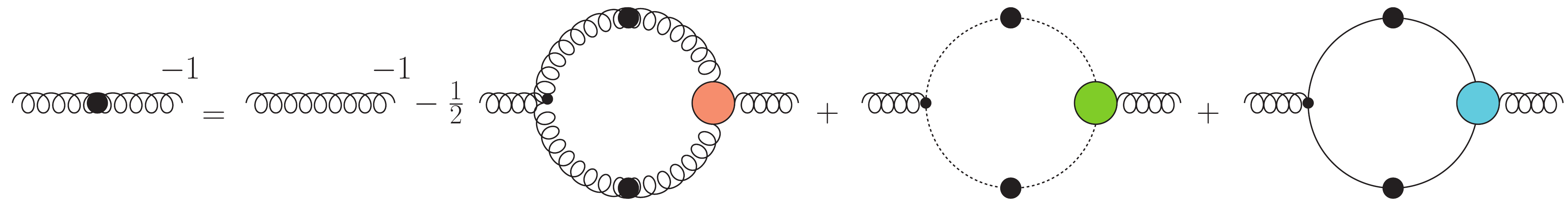}
\caption{The truncated DSE for the gluon propagator.}
\label{fig:gluon_DSE}
\end{figure*}

The gluon propagator $D^{\mu\nu}(p)$ is included self-consistently by solving the corresponding DSEs for the Yang-Mills 
system\footnote{The numerical implementation is detailed in Refs.~\cite{Hopfer:2014zna,Fischer:2003rp,Hopfer:2012ht}.}, 
where we employ the truncation scheme proposed in Refs.~\cite{Fischer:2003rp,Fischer:2002eq}.
Unquenching effects enter the gluon DSE via the quark-loop diagram as depicted in Fig.~\ref{fig:gluon_DSE}.
One obtains a closed system if the quark-gluon vertex and the three-gluon vertex is specified.
As shown later in Sec.~\ref{sec:results} the phase transition is quite insensitive to 
details of the three-gluon vertex and the main impact seems to come from different 
tensor structures immanent in the quark-gluon vertex.
Although this object was at the focus of recent investigations \cite{Williams:2014iea},
it is up to now still too ambitious to include it in a full self-consistent way due to its 
complicated multi-tensor structure.
In order to proceed we defer this desirable but also highly demanding task to future work
and model the quark-gluon vertex according to Ref.~\cite{Fischer:2003rp}.
The formal structure of the gluon DSE is given by
\begin{equation*}
 D_{\mu\nu}^{-1}(p) = Z_3 D_{0,\mu\nu}^{-1}(p) + \Pi_{\mu\nu}^{YM}(p) + \Pi_{\mu\nu}^{quark}(p),
\end{equation*}
where the gluon self-energy contribution stemming from the quark-loop reads
\begin{equation}
 \label{eq:quarkloop_SE}
 \Pi_{\mu\nu}^{quark}(p) = - g^2 \frac{N_f}{2}Z_{1F}\int\frac{d^4q}{(2\pi)^4}\, 
 tr_D\Bigl[\gamma_\mu S(q)\varGamma_\nu(q,k;p)S(k)\Bigr].
\end{equation}
We note that in general a truncated DSE system is plagued by spurious divergencies appearing in the 
kernels of the loop integrals.
Based on a UV analysis a save way to remove these unwanted 
contributions is to modify the integral kernels by constructing appropriate compensation terms, 
{\it cf.} Refs.~\cite{Fischer:2003rp,Fischer:2002eq}. For moderate flavour numbers the quark loop diagram is IR sub-leading.
Hence, a direct modification of the corresponding integral kernels is possible.
However as soon as the system approaches $N_f^{crit}$ 
the quark loop becomes IR enhanced and shows similar IR scaling
as the ghost loop. Hence, subtracting quadratic divergencies directly
from the quark loop fails if one wants to probe the chiral phase transition.
In Ref.~\cite{Hopfer:2014zna} we give several complementary methods which are able to eliminate 
these artificial contributions in a safe way and which are also used throughout.

\section{Results}
\label{sec:results}
\begin{figure}[ht!]
\center
\vspace{-0.2cm}
\subfigure{\includegraphics[width=0.49\columnwidth]{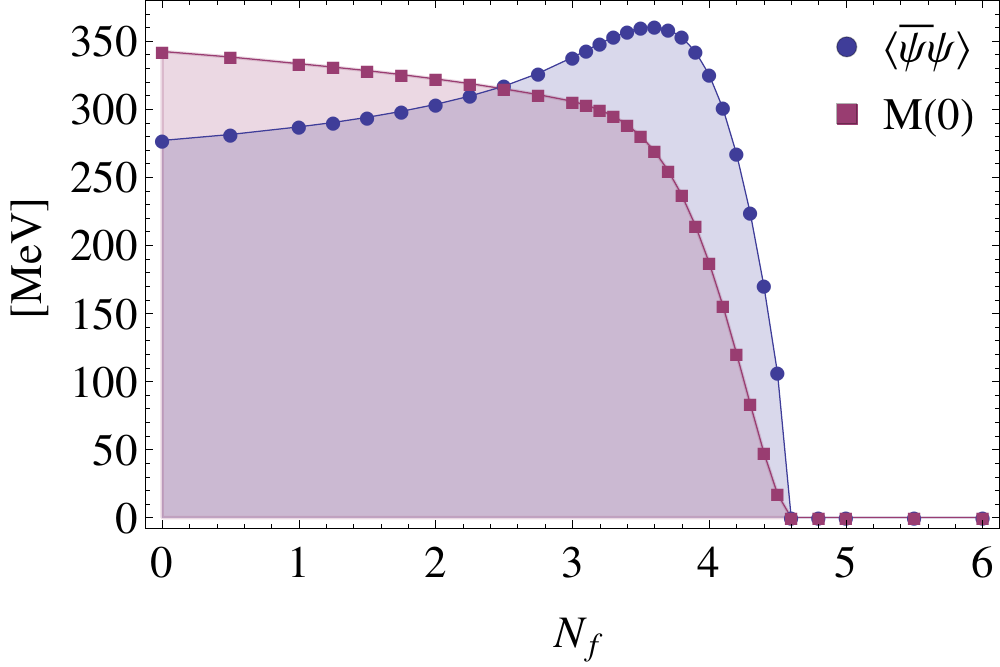}\label{fig:1BCxG2_Results_Mass}\hspace*{0.1cm}}
\subfigure{\includegraphics[width=0.49\columnwidth]{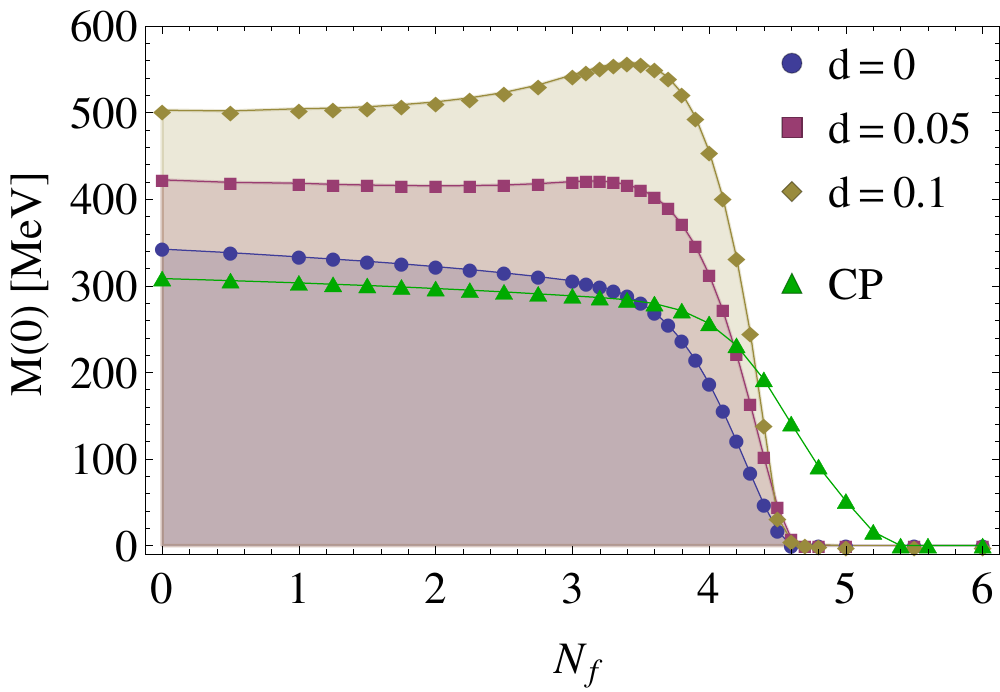}\label{fig:qgv_parameter_dependence2}} \\[-0.2cm]
\subfigure{\includegraphics[width=0.49\columnwidth]{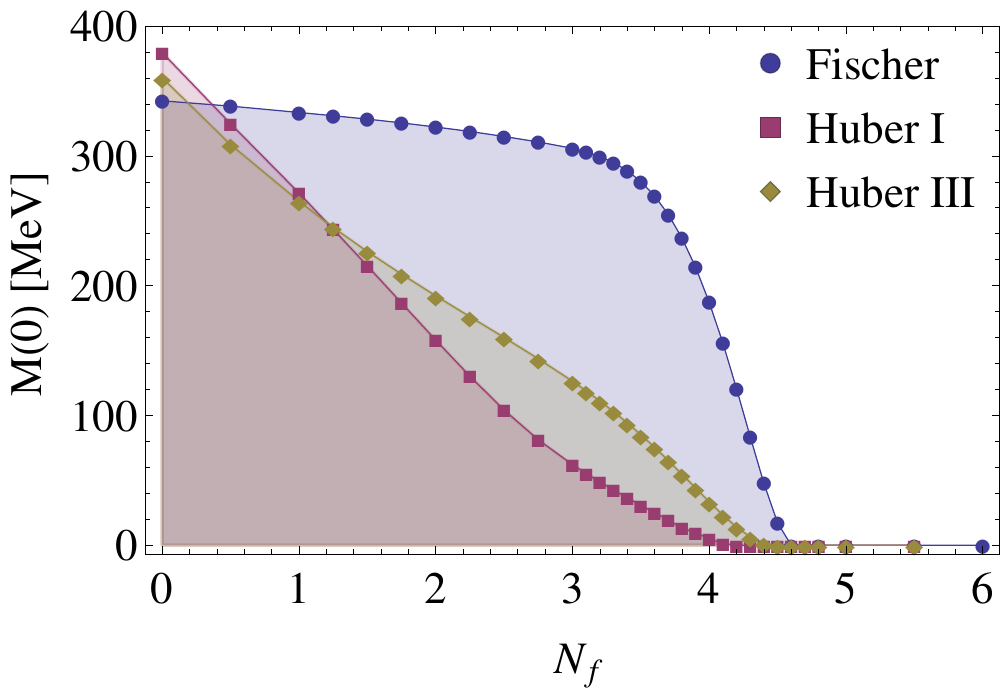}\label{fig:3gv_influence2}}\hspace{0.1cm}
\subfigure{\includegraphics[width=0.49\columnwidth]{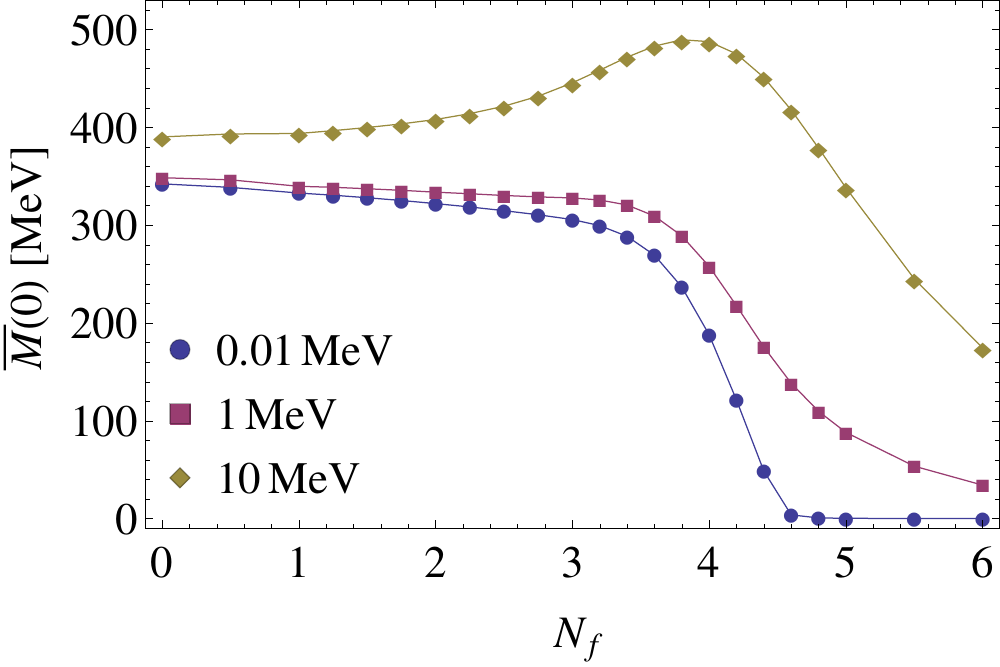}\label{fig:1BCxG2_Results_Mass_Dependence}} \\[-0.2cm]
\caption{Upper panel: The infrared quark mass function $M(p^2)$ and the chiral condensate $\langle\bar\psi\psi\rangle$ for 
different flavours $N_f$. Lines are drawn to guide the eye. 
Right: Different quark-gluon vertex models which contain \textit{only} the tree-level structure do not influence 
the phase transition.
However, \textit{additional} tensor structure increases $N_f^{crit}$.
Lower panel: The gauge-boson vertex has minor impact on the location of the transition and even tends to decrease $N_f^{crit}$.
Right: Results obtained with different bare quark masses. As expected the phase transition gets washed out.}
\end{figure}

We present results obtained from a self-consistent treatment of the DSE system on the propagator level.
For the three-point functions models are employed, where we emphasize the important role of the 
quark-gluon vertex tensor structure.
As shown in Fig.~\ref{fig:1BCxG2_Results_Mass} above $N_f^{crit} \approx 4.5$ dynamical mass is no longer generated
and the systems enters a chirally symmetric phase. 
Increasing the effective quark-gluon interaction strength using models which include \textit{only} 
the tree-level vertex structure $\gamma^\mu$ has virtually no impact on the location of the phase transition as
shown in Fig.~\ref{fig:qgv_parameter_dependence2}. We furthermore note that different models for the gauge-boson 
vertex \cite{Huber:2012kd} tend to influence $N_f^{crit}$ only slightly as detailed in Fig.~\ref{fig:3gv_influence2}.
On the other hand, including \textit{additional} tensor structure in the quark-gluon vertex increases $N_f^{crit}$ considerably.
Thus, a detailed knowledge of the quark-gluon vertex is crucial in order to give reliable predictions for $N_f^{crit}$.

\begin{figure*}[!ht]
\center
\vspace{-0.2cm}
\subfigure{\includegraphics[width=0.49\columnwidth]{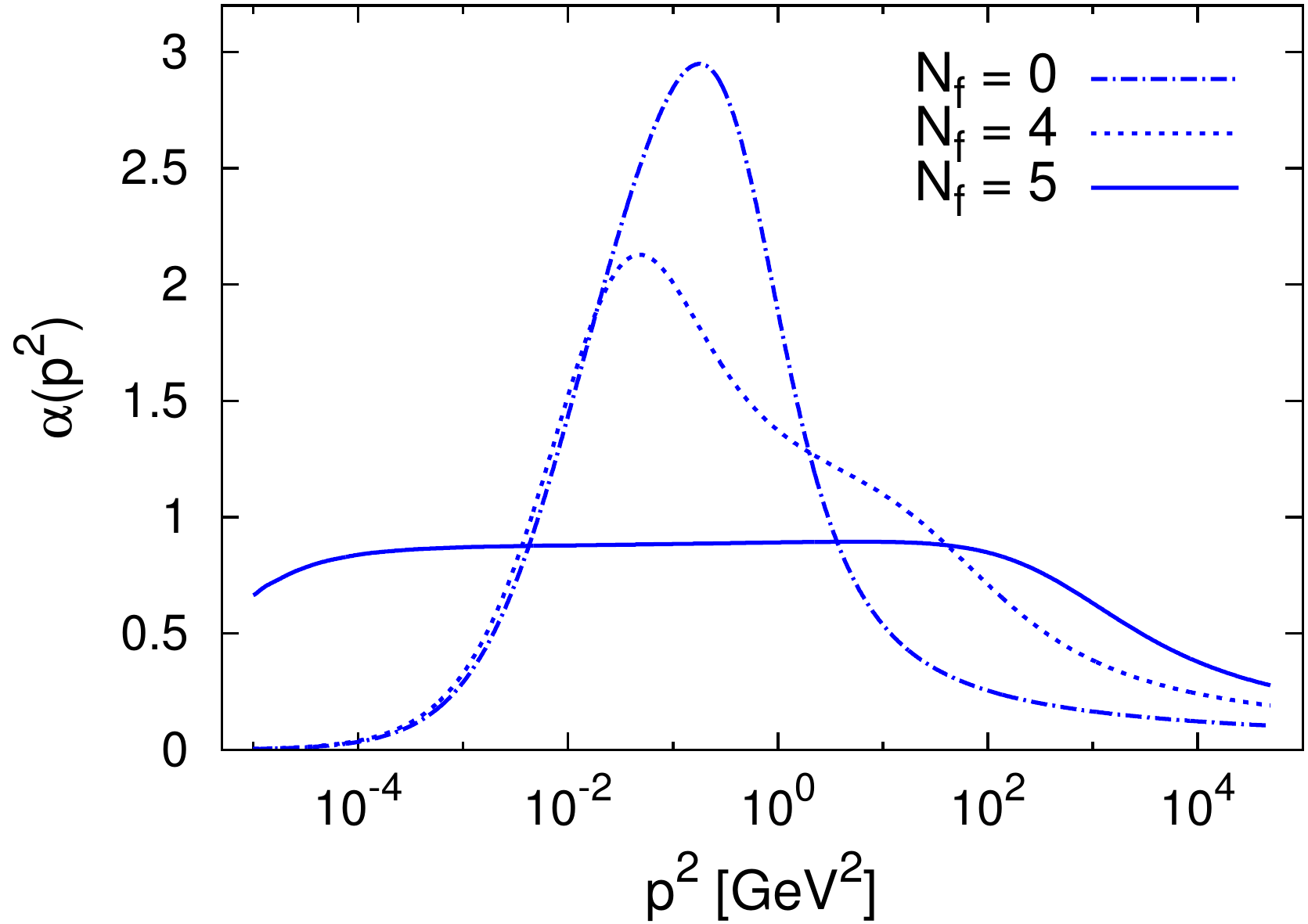}\label{fig:1BCxG2_Results_coupling}}
\subfigure{\includegraphics[width=0.49\columnwidth]{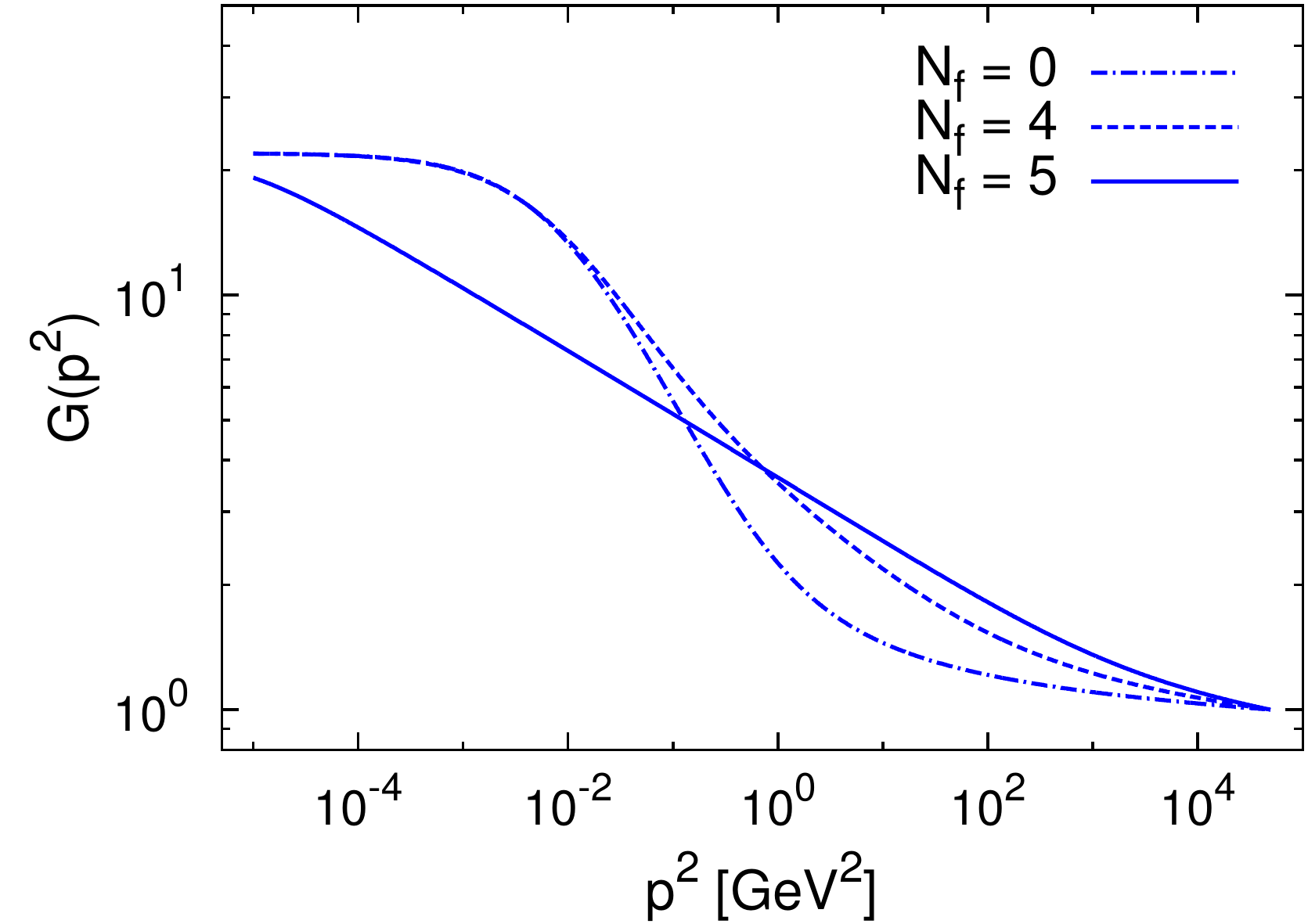}\label{fig:1BCxG2_Results_ghost}} \\[-0.1cm]
\subfigure{\includegraphics[width=0.49\columnwidth]{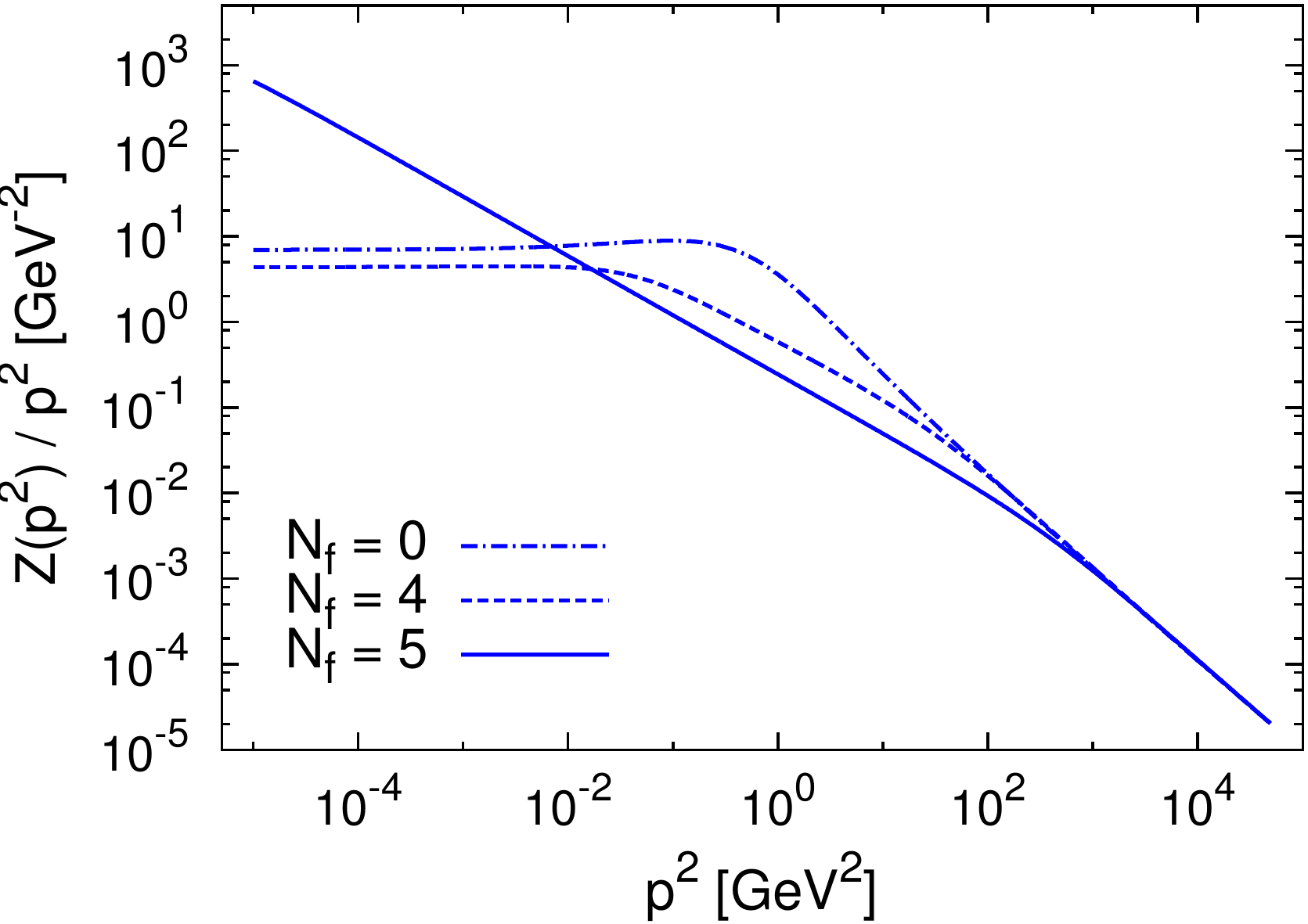}\label{fig:1BCxG2_Results_gluon}}
\subfigure{\includegraphics[width=0.49\columnwidth]{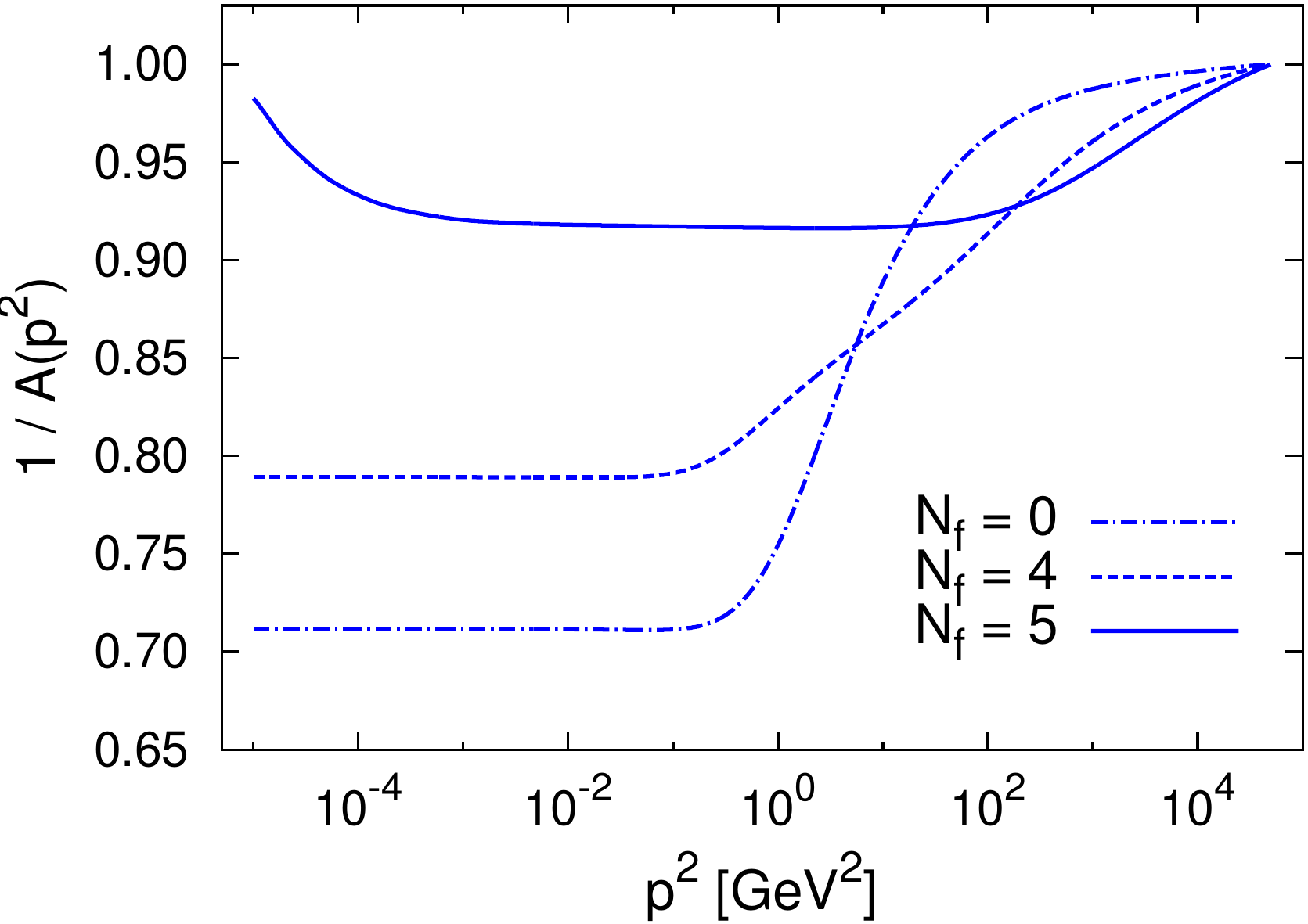}\label{fig:1BCxG2_Results_A}} \\[-0.1cm]
\caption{Results for the running coupling $\alpha(p^2)$, the ghost dressing function $G(p^2,\mu^2)$, 
the gluon propagator $Z(p^2,\mu^2)/p^2$ and the quark wave-function renormalization $A^{-1}(p^2,\mu^2)$.
In the calculations we use a perturbative renormalization scale of $\mu^2=5\times 10^4 GeV^2$.}
\label{fig:1BCxG2_Results}
\end{figure*}

The phase transition manifests itself also in a drastic change of the propagators.
In Fig.~\ref{fig:1BCxG2_Results} we present results for the non-perturbative running coupling $\alpha(p^2)=\alpha(\mu^2)Z(p^2)G^2(p^2)$,
the ghost dressing function $G(p^2)$, the gluon propagator $Z(p^2)/p^2$ and the 
inverse vector self-energy $A^{-1}(p^2)$ for different flavour values $N_f\in\{0,4,5\}$.
By increasing $N_f$ the coupling
is lowered where at $N_f\lesssim N_f^{crit}$ this lowering is significant and finally
at $N_f^{crit}$ a sudden drop occurs and a plateau is formed which develops over a wide momentum range. 
If $N_f$ is further increased the plateau is successively lowered. 
Thus, within the chirally symmetric phase a scaling relation between the Yang-Mills
propagators is established, {\it i.e.} these objects develop a power law behaviour in this momentum region
as can be seen from Fig.~\ref{fig:1BCxG2_Results_ghost} and Fig.~\ref{fig:1BCxG2_Results_gluon}.
As shown in Fig.~\ref{fig:1BCxG2_Results_A}, the quark wave-function renormalization is constant in this 
region as expected from an IR analysis.

\section{Conclusions}
\label{sec:conclusions}
We presented results from an exploratory study of large $N_f$ QCD using the Dyson-Schwinger framework in Landau gauge.
A self-consistent treatment of the corresponding DSEs on the propagator level reveals a transition to a chirally
symmetric phase for $N_f^{crit}\approx 4.5$. The non-perturbative running coupling develops a plateau in this regime, 
where, correspondingly, the propagators follow a power law indicating conformal behaviour.
The critical fermion flavour number is sensitive to details of
the quark-gluon vertex model, whereas the gauge-boson vertex seems to play a minor role.
This emphasizes the need for a more complete calculation using a full quark-gluon vertex in upcoming studies.

\vspace{-0.1cm}
\section*{Acknowledgments}
We thank Markus Huber, Axel Maas, Valentin Mader, Francesco Sannino and Milan Vujinovic for valuable discussions.
MH acknowledges support from the Doktoratskolleg ''Hadrons in Vacuum, Nuclei and Stars`` of the 
Austrian Science Fund, FWF DK W1203-N16.
\vspace{-0.1cm}


\begin{thebibliography}{99}

\bibitem{Weinberg:1979bn}
  S.~Weinberg,
  Phys.\ Rev.\ D {\bf 19} (1979) 1277;
  L.~Susskind,
  Phys.\ Rev.\ D {\bf 20} (1979) 2619;
  S.~Weinberg,
  Phys.\ Rev.\ D {\bf 13} (1976) 974;
  S.~Dimopoulos and L.~Susskind,
  Nucl.\ Phys.\ B {\bf 155} (1979) 237;
  E.~Eichten and K.~D.~Lane,
  Phys.\ Lett.\ B {\bf 90} (1980) 125;
  E.~Farhi and L.~Susskind,
  Phys.\ Rept.\  {\bf 74} (1981) 277.

\bibitem{Sannino:2009za}
  F.~Sannino,
  Acta Phys.\ Polon.\ B {\bf 40} (2009) 3533
  [arXiv:0911.0931 [hep-ph]].


\bibitem{Bashir:2013zha}
  A.~Bashir, A.~Raya and J.~Rodriguez-Quintero,
  Phys.\ Rev.\ D {\bf 88} (2013) 054003
  [arXiv:1302.5829 [hep-ph]].

\bibitem{Hopfer:2014zna}
  M.~Hopfer, C.~S.~Fischer and R.~Alkofer,
  arXiv:1405.7031 [hep-ph].

 \bibitem{Fischer:2003rp}
  C.~S.~Fischer and R.~Alkofer,
  Phys.\ Rev.\ D {\bf 67} (2003) 094020
  [hep-ph/0301094];
  C.~S.~Fischer, PhD thesis [hep-ph/0304233];
  C.~S.~Fischer, P.~Watson and W.~Cassing,
  Phys.\ Rev.\ D {\bf 72} (2005) 094025
  [hep-ph/0509213].

\bibitem{Hopfer:2012ht}
  M.~Hopfer, R.~Alkofer and G.~Haase,
  Comput.\ Phys.\ Commun.\  {\bf 184} (2013) 1183
  [arXiv:1206.1779 [hep-ph]].

\bibitem{Fischer:2002eq}
  C.~S.~Fischer, R.~Alkofer and H.~Reinhardt,
  Phys.\ Rev.\ D {\bf 65} (2002) 094008
  [hep-ph/0202195];

\bibitem{Williams:2014iea}
  R.~Williams,
  arXiv:1404.2545;
  M.~Hopfer, A.~Windisch and R.~Alkofer,
  PoS CONFINEMENTX {\bf } (2013) 073
  [arXiv:1301.3672];
  R.~Alkofer, C.~S.~Fischer, F.~J.~Llanes-Estrada and K.~Schwenzer,
  Annals Phys.\  {\bf 324} (2009) 106
  [arXiv:0804.3042].

\bibitem{Huber:2012kd}
  M.~Q.~Huber and L.~von Smekal,
  JHEP {\bf 1304} (2013) 149
  [hep-th/1211.6092];
 M.~Q.~Huber and L.~von Smekal,
  PoS CONFINEMENTX {\bf } (2013) 062
  [hep-th/1301.3080].

\end{thebibliography}
\end{document}